\documentclass[12pt]{article}

\title{Non stationary nucleation: the model with minimal environment}
\author{Victor Kurasov}
\date{St.Petersburg State University}

\begin{document}
\maketitle

\begin{abstract}

A new model to calculate the rate of nucleation is formulated. This model is based on the classical nucleation theory but considers also vapor depletion around the formed embryo. As the result the free energy has to be recalculated which brings a new expression for the nucleation rate.

\end{abstract}

\section*{Introduction}

A typical example of the first order phase transitions is the  condensation of a supersaturated vapor in a state of liquid droplets. It is wide spread in nature and became a model for theoretical investigations. Namely in this case the most advanced models of the phenomena were presented. The start of systematic investigations was given by the famous Wilson's chamber already at the end of the 19-th century \cite{Wilson}. Despite the long history there are still some unresolved problems in coincidence of the theoretical description with experimental results. Experimental technique is rather accurate, theoretical models are sophisticated but the measured rate of nucleation can differ many times from the theoretical predictions.

Creation of the so-called "classical theory of nucleation" by Becker and  Doering \cite{BD}, Volmer and Weber \cite{Volmer}, Zeldovich \cite{Zeldovich}, Frenkel \cite{Frenkel} gives the expression for the stationary rate of nucleation which is rather evident but contains the free energy of the critical embryo as a parameter. The last value has to be derived in frames of some model for the critical droplet. The reliable model is the capillary approximation which applies the standard conceptions of thermodynamic to the droplet.

The classical theory of nucleation was mainly completed at 40-ties but experimental measurements did not confirm predictions of the theory - the experimental values of the rate of nucleation can many times differ from theoretically given values. The most radical reconsideration of the theoretical description was given by Lothe and Pound \cite{LP} who suggested that inside the droplet a special account of rotation degrees of freedom has to be made. This initiated a wide discussion, many arguments were presented but there is no strong arguments to state that all rotational degrees of freedom have to be accounted in a special way. The only evident modification is to make a special account of rotational degrees of freedom for the embryo as a whole object. It was done by H. Reiss, J. L. Katz, E.R. Cohen\cite{RKC} but this modification can not bring the theory in coincidence with experiment. Honestly speaking one has to declare that the theory of Lothe and Pound is also far from experimental results for some substances meanwhile it gives some promising results for the water condensation. After numerous strikes and discussions in 60-ies and 70-ies the problem of discrepancy between theoretical and experimental results has not yet disappeared.

One has to stress that the difficulty is not in the form of the rate of nucleation which is
$$
J_{theor} = N_{eq} Z \exp(-F_c)
$$
where $F_c$ is a free energy of the critical embryo, $N_{eq}$ is a normalizing factor of the equilibrium distribution, $Z$ is so-called Zeldovich factor. The main technical efforts of the classical nucleation theory were intended to  determine  $Z$. 

But the problem is also not in determination of $Z$ which is evidently proportional to kinetic coefficient $W^+$ (which is the inverse mean time between collisions of the critical embryo with molecules of vapor) and to the condensation coefficient $\alpha_c$.  The typical error in  $Z$ under the rough estimation  results in some essential value, but not in order of this magnitude. So, the error which is the source of discrepancy between theoretical and experimental results can be made in $N_{eq}$ or in $F_c$. Since these values appear together in the value of the equilibrium distribution  $N_{eq} \exp(-F_c)/ \exp(1)$ which is a boundary condition at the left boundary of the near-critical region one can not separate them. Historically it is preferable to speak about the error in $F_c$ regarding $N_{eq}$ as some smooth value like the number of monomers in a system.

\section*{Formulation of the problem}

One can say that the theoretical models are not so sophisticated and their further development can bring the theory in coincidence with experiment. Creation and development of multi-component theory show that the flow will go mainly through the saddle point and the free energy $F_s$ at the saddle point can be only lower than the free energy of the critical cluster calculated in the one-component theory. The reasons for the multi-component consideration can be very various, for example, the existence of some compact complexes of molecules inside the embryos. But it is clear that any more sophisticated consideration can lead only to the  decrease of the free energy of the critical embryo and, thus, to the increase of the nucleation rate. But the the experimental rate of nucleation $J_{exp}$ can be less than the theoretical rate $J_{theor}$ for some substances and the reconsideration of the theory can only diminish the last value. So, in the situation
$$
\frac{J_{exp}}{J_{theor}} \ll 1
$$
 the new more sophisticated models of the embryo can only enlarge the discrepancy.

Certainly, one can attribute the discrepancy between theoretical and experimental values to the coefficient of condensation $\alpha_c$. But it is hard to believe that the coefficient of condensation can be like $10^{-5\div -7}$ which is necessary to ensure the coincidence. This value does not correspond to the values $10*{0 \div -2}$ of the condensation coefficient observed in experiments for the condensation in the bulk liquid. Really, the small droplet and the bulk liquid are different objects but one and the same molecular content of these objects makes it difficult to believe in such difference.

Another possibility is to attribute this discrepancy to the microscopic corrections  for the free energy. These corrections are very complex functions and are quite different for different substances. But here appeared another problem. The characteristic weight of these corrections for big embryos has to be relatively small. So, there has to be a sharp functional dependence of the ratio of the rates without and with essential corrections  on the size of the critical embryo. But the real situation shows a smooth value for the ratio of theoretical and experimental rates of nucleation.There is no evidence why in decomposition on inverse radius of embryo for microscopic corrections the third term has to attain an enormously big value.

As the result we see that there has to be some another mechanism which is responsible for the decrease of the nucleation rate in comparison with the classical consideration. One of the possible answers is that the free energy which stands in the main term for the theoretical rate
$$
J_{theor} \sim \exp(-F_c)
$$
is not a true equilibrium free energy. The minimal work of the embryo formation can be attained only at the infinitely slow process of the embryo formation and here the situation is not so slow as it is necessary to apply the classical expression for the free energy
$$
F(\nu) = a \nu^{2/3} - b \nu
$$
of the embryo with $\nu$ molecules where $a$ is the normalized surface tension and $b$ is the difference  of the bulk chemical potentials.

Here a careful reader has to argue because in the classical theory of nucleation there exists a solid justification to apply the thermodynamic expression for the free energy. We shall repeat here this justification.

The embryos are so small and the liquid is so dense in comparison with vapor that the time of the diffusive and thermal relaxation is very small in comparison with  the mean time between sequential collisions of the embryo with molecules of substance in vapor. The time of relaxation can be estimated as
$$
t_{rel} = R_d^2 / 4 D_l
$$
where
$R_d$ is the radius of the embryo and $D_l$ is the coefficient of diffusion in liquid. The time between sequential collisions is $W_+^{-1}$ where $W_+$ is the direct flow of the molecules on the embryo. In the free molecular regime it can be estimated as
$$
W_+ \approx \frac{1}{4} v_t n S
$$
where
$v_t$ is the mean thermal velocity of molecules in vapor, $n$ is the density of the particles number, $S$ is the square of the embryo. One can check the strong inequality
$$
t_{rel} \ll W_+^{-1}
$$
and see that it really takes place.

 The point of view adopted in this theory states that there is no strict connection between this justification and the possibility to write the free energy only for the embryo. The embryo at such big times as the time of the while evolution can perturb the environment meanwhile it is a quasi-isolated object at the relatively small times. All adiabatic approaches deal with such situations. But nobody states that it is necessary to forbid the change with environment and to use only the static characteristics.

So, here we shall consider the environment of the embryo and require that the embryo has to be in the stable state being embedded in this environment.

\section*{Equilibrium environment}
There exists one argument which is rather unpleasant for the formulated non-stationary theories. Really, after the precritical embryo is dissolved the vapor environment is still perturbed. Now there is an excess of the molecules in vapor near the place where the embryo was   previously formed. So, the story of this fluctuation continues. In some cases it will be completely dissolved, in some cases it will give a new life for a new embryo. So, we see that we come to a level of fluctuations in vapor.We need to consider not a single embryo but a fluctuation of molecules density where there is an embryo and this embryo is in equilibrium conditions.

The main supposition used here will be a requirement to build the fluctuation of a minimal size corresponding to a possibility to have an equilibrium embryo. Later this supposition will be explained and justified.

\section*{Construction of the theory}

Considering the fluctuation containing the liquid phase and the vapor phase we need to calculate the minimal work to form such a fluctuation. This work has to correspond to the equilibrium process and, thus, the fluctuation has to be an equilibrium one. Then, we turn to the question what fluctuation will be equilibrium one.

Consider the system of $N$ molecules and an embryo of $\nu$ molecules inside it. At this very moment the number $N$ is unknown but later we shall express it through $\nu$. The substance balance gives
$$
N=\nu +q
$$
where $q$ is the number of molecules in vapor. The density of the number of molecules in vapor phase is
$$
n = q/V
$$
where $V$ is the volume of the fluctuation. If the initial density $n_0$ of the vapor molecules is known then
$$
V=N/n_0
$$
Then in the expression $F=a\nu^{2/3}-b\nu$ one has to recalculate the difference in chemical potentials $b$ on the base of $n$ instead of $n_0$. Namely, we have to substitute $n$ instead of $n_0$ in the expression
$$
b = \ln(n_0/n_{\infty})
$$
where $n_{\infty}$ is the molecular number density in the saturated vapor.
Earlier this expression was
$$
b=\ln(\zeta+1) \ \ \ \zeta=(n_0-n_{\infty})/n_{\infty}
$$
now it is
$$
b=\ln((n-n_{\infty})/n_{\infty}+1)=\ln(((N-\nu)n_0/N-n_{\infty})/n_{\infty}+1)
$$

Then
$$
F=a\nu^{2/3} - \nu \ln(\zeta+1) - \nu \ln(1-\frac{\nu}{N})
$$

Since the fluctuation has to be the equilibrium one to ensure the minimal work of formation we seek the local minimum of the free energy $F$ and the argument of this minimum has to be the true number $\nu_e$ of the molecules inside the embryo
$$
\nu_e = arg( min_{\nu} F(\nu, N))
$$
So, for given $\nu_e$ we get the value of $N$ which provides the minimum of $F$ namely at $\nu_e$. It gives the dependence $N=N(\nu_e)$. The value $\nu_e$ is the true value of $\nu$ to be used if we want to calculate the free energy of the embryo in the stable fluctuation. So, we get then the value $F_e$ as the value of the local minimum of the free energy, i.e.
$$
F_e = F(\nu_e, N(\nu))
$$
This value will be the true value of the free energy of the embryo containing $\nu$ molecules in frames of this model which will be called "the model with the minimal environment". We shall keep the name $F_{eq}$ for the free energy calculated in this model

Expression for $dF/d\nu$  looks like
$$
dF/d\nu=\frac{2}{3}a\nu^{-1/3} -  \ln(\zeta+1) -  \ln(1-\frac{\nu}{N}) + \frac{\nu}{(1-\frac{\nu}{N})N}
$$
and the equation of extremum
$$
\frac{2}{3}a\nu^{-1/3} = \ln(\zeta+1) +  \ln(1-\frac{\nu}{N}) - \frac{\nu}{(1-\frac{\nu}{N})N}
$$
is very simple but can not be solved analytically. So, it is not clear whether the free energy $F(\nu, N)$ really have necessary minimum (the minimum at the origin does not match). So, it is necessary to see the properties of the free energy in this model.

Instead of analyzing the dependence $N(\nu_e)$ we shall consider dependence $\nu_e(N)$ and we see immediately that when $N$ goes to infinity then $\nu_e$ also goes to infinity and $F_e$ will attain a big negative value. This shows that if for some $\nu$ we can ensure the minimum, i.e. $\nu=\nu_e$ then one can provide minimum for all $\nu$ greater than the mentioned one. The problem is for small $\nu$ one can not get $N$ providing $\nu_e$. One can see it from considerations below.

We shall only diminish the action of correction term if we linearize the logarithm. It is so because
$\nu/N$ is positive and $\ln$ is a convex function.  Then
$$
F=a\nu^{2/3} - \nu \ln(\zeta+1) +  \frac{\nu^2}{N}
$$
For this function one can easily calculate the second derivative
$$
d^2 F / d\nu^2 = - (2a/9)\nu^{-4/3} + \frac{2}{N}
$$
Then for the transition to the absence of the threshold we have
$$
\nu = (Na/9)^{3/4}
$$
We see that the critical (minimal) value for $N$ ensuring a threshold  is
$$
N_c = 9 \nu^{4/3} / a
$$
which is a big value and the argument of decomposition $\nu/N$ is small at $\nu \gg 1$ which a natural requirement for thermodynamics. Then the decomposition is quite possible.

For the critical value of $N_c$ we have
$$
F = a\nu^{2/3} - \nu \ln(\zeta+1) +  \frac{\nu^2a}{9\nu^{4/3} }
$$
or
$$
F = a(1+1/9)\nu^{2/3} - \nu \ln(\zeta+1)
$$
Really, we need only the barrier value of the free energy which corresponds to $N=N_c$.

So, we come to the situation with a redefined surface tension
$$
a \rightarrow a(1+1/9)
$$
The surface tension grows $1+1/9$ times.

This conclusion is very important. On one hand it allows to formulate a new variant of the Gibb's theorem about the proportionality of the free energy of the critical cluster to the part of the surface energy. Now all constructions will be the same but only the quantity of one third will be redefined. On the other hand it explains numerous conclusions in experimental papers that everything would be good with theoretical predictions if the surface tensions is redefined. Now the reason is obvious.

 \section*{The justification of the model}

Now it is clear why it was necessary to require the fastest creation of the critical embryo or the fastest transition over the activation barrier. We see that the precritical cluster is very unstable formation (there is even no stable environment). When the molecules will be spread the volume with the linear size like
$$
R_w = (\nu v_v)^{1/3}
$$
there will be no trace of the embryo. Here $v_v$ is the characteristic volume for one molecule in the saturated vapor. The characteristic time can be estimated as
$$
t_w = R_w^2/(4D)
$$
where $D$ is a coefficient of diffusion in vapor.

Here we take the longest estimate corresponding to the diffusion relaxation while the disappearance of the embryo can occur faster.

We have seen that to make the stable configuration it is necessary to spread the volume of the influence up to the linear size
$$
R_q = (N_c v_v)^{1/3}
$$
One can see that
$$
R_q/R_w \sim \nu^{1/9} \gg 1
$$
The characteristic time here can be estimated as
$$
t_q = R_q^2/(4D)
$$
Since
$$
t_q/ t_w \sim \nu^{2/9} \gg 1
$$
it is necessary to gather the embryo as soon as possible and, thus, to gather the molecules from the minimal volume ensuring the stability of the fluctuation. Namely this was done in this model.

\section*{The rate of nucleation}

Since $Z$ does not strongly depend on supersaturation one can take it in expression for the rate of nucleation being calculated in frames of the old classical theory. This helps to ignore the absence of explicit barrier at the situation with minimal environment. The value of the normalizing factor $N_{eq}$ also can be taken from the old variant of the nucleation theory. One has to stress  that here one can use as the reference theory all already existing approaches including Lothe and Pound approach \cite{LP}, Reiss, Kats, Cohen approach \cite{RKC}, etc. All already formulated approaches have nothing in common with depletion and the finite time of formation of the critical embryo  and deals with another effects. So, they can be really combined with this approach without any difficulties in justification. Technical difficulties are also absent.

One has to  stress that here it is possible to avoid to linearize the  logarithm of the new (correction) term in the expression for the free energy. Expressions will be slightly more complex but this approach will give the better accuracy. This improvement is important because the small relative shift in the height of barrier will radically change the value of the nucleation rate. Since the formulas without this simplification are not explicit ones it is preferably to use them directly in numerical calculations. The scheme of approach remains the same.

Here the main attention is payed to the period of formation of the embryos. It can seem that this period is not too important because in the process of the global condensation one can see the universal features \cite{Mygolden}, \cite{Myuniversal}, \cite{Myperturbative}. The final asymptotic of over-condensation (coalescense) is universal \cite{LS} also. But the recent investigations of the over-condensation \cite{Mylate}, \cite{Mylate2} show that the initial data are very important in the final stages of the process and can determine the final behavior even qualitatively.

  \end{document}